\documentclass[a4paper,fleqn,usenatbib]{mnras}

\usepackage[T1]{fontenc}
\usepackage{ae,aecompl}

\usepackage{graphicx}	
\usepackage{amsmath}	
\usepackage{amssymb}	
\usepackage{type1ec}
\usepackage{type1cm}
\usepackage{hyperref}

\newcommand{\comment}[1]{}

\title[Radio Counterparts of SLSNe]
{Radio Emission from Embryonic Super-Luminous Supernova Remnants}
\author[Omand et al.]
{Conor M. B. Omand$^{1}$\thanks{E-mail:omand@utap.phys.s.u-tokyo.ac.jp}, 
Kazumi Kashiyama$^{1,2}$, 
and Kohta Murase$^{3,4,5,6}$\\
$^{1}$Department of Physics, School of Science, the University of Tokyo, Tokyo 113-0033, Japan\\
$^{2}$Research Center for the Early Universe, the University of Tokyo, Tokyo 113-0033, Japan;\\
$^{3}$Department of Physics, The Pennsylvania State University, University Park, PA 16802, USA\\
$^{4}$Department of Astronomy \& Astrophysics, The Pennsylvania State University, University Park, PA 16802, USA\\
$^{5}$Center for Particle and Gravitational Astrophysics, The Pennsylvania State University, University Park, PA 16802, USA\\
$^{6}$Yukawa Institute for Theoretical Physics, Kyoto University, Kyoto 606-8502, Japan}

\date{Accepted XXX. Received YYY; in original form ZZZ}

\pubyear{2017}

\begin{document}
\label{firstpage}
\pagerange{\pageref{firstpage}--\pageref{lastpage}}
\maketitle

\begin{abstract}
It has been widely argued that Type-I super-luminous supernovae (SLSNe-I) are driven by powerful central engines with a long-lasting energy injection after the core-collapse of massive progenitors. 
One of the popular hypotheses is that the hidden engines are fast-rotating pulsars with a magnetic field of $B\sim{10}^{13}-{10}^{15}$~G. 
Murase, Kashiyama \& Meszaros (2016) proposed that quasi-steady radio/submm emission from non-thermal electron-positron pairs in nascent pulsar wind nebulae can be used as a relevant counterpart of such pulsar-driven supernovae (SNe). 
In this work, focusing on the nascent SLSN-I remnants, we examine constraints that can be placed by radio emission. 
We show that the Atacama Large Millimeter/submillimetre Array (ALMA) can detect the radio nebula from SNe at $D_{\rm L} \sim 1 \ \rm Gpc$ in a few years after the explosion, 
while the Jansky Very Large Array (VLA) can also detect the counterpart in a few decades.
The proposed radio followup observation could solve the parameter degeneracy in the pulsar-driven SN model for optical/UV light curves, 
and could also give us clues to young neutron star scenarios for SLSNe-I and fast radio bursts. 
\end{abstract}


\section{Introduction}
Super-luminous supernovae~(SLSNe) are extremely rare, but they are the most luminous optical/UV transients associated with massive stellar deaths~\citep[e.g.,][]{Gal-Yam12}.
They are divided into at least two groups, depending on the presence of hydrogen signatures in the observed supernova (SN) spectra. 
Those with hydrogen (Type-II SLSNe; SLSN-II) are likely powered by a circumstellar shock between the SN ejecta and massive hydrogen-rich envelope~\citep[e.g.,][]{Smith_McCray_07,Chevalier_Irwin_11},   
while those without (Type-I SLSNe; SLSN-I) are believed to originate from massive progenitors such as Wolf-Rayet stars, and are probably powered by a central engine, either a fast spinning newborn pulsar~\citep[e.g.,][]{Woosley_2010,kb10} or a black-hole accretion disk system~\citep{Dexter_Kasen_13}. 

The pulsar-driven scenario for SLSNe-I has attracted attention in the context of explaining the diversity of cosmic explosions that are related to deaths of massive stars 
\citep[e.g.,][]{2010ApJ...724L..16P,qui+11,Inserra_et_al_2013,2014MNRAS.444.2096N,2016ApJ...828L..18N,Metzger_et_al_15,Wang_et_al_2015,Dai_et_al_2016}
Millisecond-spinning strongly-magnetized neutron stars (NSs) have been implemented as a central engine of gamma-ray bursts~(GRB) and broad-line Type Ibc SNe (or hypernovae)~\citep[e.g.,][]{Thompson_et_al_2004}.  
The GRB-SN connection has been extended to include more luminous SN classes, in particular after the discovery of an SLSN-like counterpart of an ultra-long GRB~\citep{Greiner_et_al_15,Metzger_et_al_15}. 
On the other hand, stripped-envelope SNe, including ordinary Type Ibc SNe, may also be accompanied by strongly magnetized pulsars. 
The magnetar formation rate is estimated to be $\sim10$\% of the core-collapse SN rate \citep{kk08}, 
and Galactic magnetars that are slowly rotating for now could be explained by the formation of strongly magnetized NSs with a a relatively slow initial spin period of $\gtrsim 10 \ \rm ms$~\citep{Kashiyama+16}.   
It is important to understand the diversity of stripped-envelope SNe (in particular SLSNe-I, hypernovae, and SNe Ibc), which may enable us to gain insights about formation mechanisms of fast-rotating pulsars or magnetars.  

What is the smoking gun of a nascent pulsar embedded in the SN ejecta? 
In general, the SN emission can be powered by radio-active nuclei, and we often suffer from the parameter degeneracy 
when we only use information on light curves~\citep[e.g.,][]{Wang_et_al_2015,Kashiyama+16}. In addition, the energy injection does not have to be caused by the pulsar itself, either~\citep{Dexter_Kasen_13}. 
Non-thermal signatures can be used as a useful probe of the hidden compact remnants. Indeed, Galactic pulsar wind nebula (PWNe) are established as efficient accelerators of electrons and positrons~\citep[e.g.,][]{Gaensler_et_al_06,Tanaka_Takahara_10}. If the spin-down energy is efficiently converted into radiation (as required in the pulsar-driven SN model), it is natural to expect that they are powerful emitters of X-rays and gamma-rays. 
The theory has predicted that synchrotron X-rays from leptons accelerated in PWNe provide the most promising signals, and both soft X-rays~\citep{Metzger_et_al_2013} and hard X-rays~\citep{Kashiyama+16}, are detectable by the current X-ray telescopes such as {\it Swift} and {\it NuSTAR}. 
Searches for X-ray counterparts of SNe have also been done in the past, and some tentative candidates were reported~\citep{Perna_Stella_2004,per+08,Margutti_et_al_17}. 
However, the detectability of X-rays depends on plasma properties of the SN ejecta (especially in the soft X-ray range), and the present results from X-ray measurements are not strongly constraining. On the other hand, high-energy gamma-rays, which can be produced by the inverse-Compton scattering with thermal SN photons~\citep{kot+13,2015ApJ...805...82M}, provide a more direct probe of the hidden pulsar, but detecting such signals is generally more challenging.  

Radio emission serves as an alternative probe of non-thermal acceleration in nascent PWNe embedded in the SN ejecta. \cite{MKM16} calculated synchrotron emission from young PWNe as well as various effects of gamma-ray and radio attenuation, and even considered the connection to fast radio bursts (FRBs).  \cite{MKM16} showed that quasi-steady radio/submm emission from PWNe associated with pulsar-driven SN remnants is detectable with current facilities such as he Karl G. Jansky Very Large Array (VLA) and  Atacama Large Millimeter/submillimetre Array (ALMA). Since young NSs are also thought to be the candidate progenitors of FRBs, follow-up searches for radio/submm counterparts of FRBs as pulsar-driven SN remnants were proposed to investigate their links. Interestingly, the recent observations of the repeating FRB 121102 led to the discovery of its host galaxy, and a persistent radio counterpart was seen by VLA and the European VLBI Network~\citep{Chatterjee_et_al_17,Marcote_et_al_17,Tendulkar_et_al_17}.
The identified host of FRB 121102 is a star-forming dwarf galaxy with a relatively low metallicity, which is consistent with observed hosts of SLSN-I~\citep{Chatterjee_et_al_17,Tendulkar_et_al_17,2017arXiv170102370M}.
Also the persistent radio counterpart of FRB 121102 is broadly consistent with PWNe emission from a pulsar-driven SLSN remnant a few decades old~\citep{MKM16,2017arXiv170102370M,2017ApJ...839L...3K}. 
These motivate radio follow-up observations of known candidates of pulsar-driven SNe, targeting the radio PWNe, and SLSNe-I are among the most interesting objects. 

In this work, we study quasi-steady radio emission from embryonic SLSN remnants for up to a few decades after the explosion. 
First, we select some of the known brightest SLSNe-I and fit the light curves using the pulsar-driven SN model~(Sec. \ref{sec:SN}). 
Then, using the obtained model parameters, i.e., initial spin period and magnetic field strength of the NS, and ejecta mass, 
we consistently calculate radio emission from their nascent PWNe~(Sec. \ref{sec:radio}).
We show the detectability with VLA and ALMA and discuss possible constraints on the pulsar-driven scenario.  

\section{Pulsar-driven Super-luminous Supernovae}\label{sec:SN}

In the pulsar-driven scenario, the energy source is the rotational energy of a newborn NS, 
which is extracted as a strongly magnetized relativistic wind. 
The pulsar wind is injected into the SN ejecta, driving forward and reverse shocks. 
The reverse shock region is often called a nebula or PWNe, where the pulsar wind is dissipated and electrons and positrons are accelerated up to ultra-relativistic energies. Non-thermal emission from PWNe has been studied for many years 
~\citep[e.g.,][]{Gaensler_et_al_06,Tanaka_Takahara_10}. From the modeling of Galactic PWNe, it is known that most of the spin-down energy is used for particle acceleration, and accelerated electrons and positrons lose their energies via not only adiabatic losses but also synchrotron and inverse Compton emission, which constitute a broad-band spectrum from radio to gamma-ray bands. 

It is natural to expect that the dissipation of pulsar winds and resulting particle acceleration occur in the early states of PWNe. However, the non-thermal PWNe emission is initially completely down-scattered or absorbed, and then it diffuses out from the SN ejecta as thermal radiation, being observed as SN emission in the optical band~\citep{Woosley_2010,kb10}. Physical parameters of the newborn pulsar, namely spin period and dipole magnetic field strength, can be inferred from the SN light curve~\citep[e.g.,][]{Pastorello_et_al_2010,Inserra_et_al_2013,nic+13}. 
Only in the late stages, radio emission can escape from the PWNe and dense SN ejecta without significant attenuation. In the context of pulsar-driven SNe including SLSNe, \cite{MKM16} concluded that radio emission may escape at $\gtrsim 10$~yr while the submm emission may escape at $\sim{\rm a~few}$~yr, which is consistent with more recent results by \cite{2017arXiv170102370M}.

\subsection{Supernova Samples}\label{sec:opdata}

\begin{table*}
\caption{Properties of the SLSNe selected for this study. } 
\begin{center}
\begin{tabular}{|c|c |c|c| c|c|c|} \hline
Name & RA & Dec & z & $D_L$ (Gpc) & Band & References \\ \hline 
iPTF13ajg & 16:39:03.95 & +37:01:38.4 & 0.7403 & 4.6736 & R & \citep{2014ApJ...797...24V} \\
SN2012il &  09:46:12.91 & +19:50:28.7 & 0.175 & 0.8686 & r & \citep{Inserra_et_al_2013} \\
SN2013dg &  13:18:41.35 & -07:04:43.0 & 0.1918 & 0.9615 & r & \citep{2014MNRAS.444.2096N} \\
SN2010gx &  11:25:46.71 & -08:49:41.4 & 0.2297 & 1.1766 & r' & \citep{2010ApJ...724L..16P}\\
SN2011ke &  13:50:57.77 & +26:16:42.8 & 0.1428 & 0.6950 & V & \citep{Inserra_et_al_2013} \\
SN2015bn &  11:33:41.57 & +00:43:32.2 & 0.1136 & 0.5427 & V & \citep{2016ApJ...828L..18N}  \\  \hline 
\end{tabular}
\label{tbl:snprop}
\end{center}
\end{table*}

We retrieve SLSN data from the Open Supernova catalog \footnote{https://sne.space/} \citep{2017ApJ...835...64G}.
We select SN samples where there are data points on both sides of the light curve maximum in a single band.  
There are six SNe that fit this criteria; their properties are summarized in Table~\ref{tbl:snprop} and the observed light curves are shown in Fig. \ref{fig:snopmods}.

\subsection{Modeling of Light Curves}
To calculate optical light curves, we adopt the model from \cite{Kashiyama+16}.  This model allows us to numerically calculate broadband emission, which may account for electromagnetic and gravitational wave emission due to the spin-down activity of a pulsar, acceleration of the ejecta through a magnetized wind, and radioactive decay of $^{56}$Ni and $^{56}$Co.  The model is calibrated, without a pulsar engine, against the Arnett model \citep{1982ApJ...253..785A} and thus gives similar results \citep[see][Figure 16]{Kashiyama+16}. Thermalization of non-thermal photons is also taken into account, by approximating results of \cite{2015ApJ...805...82M}, which takes into account simplified particle creation and pair cascades. Radio and submm emission is calculated based on \cite{MKM16}, which proposed the importance of radio and submm emission as a probe of pulsar-driven SNe.
Three parameters that we vary are the initial spin period $P$ and the magnetic field $B_{13} = B$/($10^{13}$ G) of the NS, and the mass $M_{\rm ej}$ of the SN ejecta.  
Other parameters in the model are the ejected nickel mass $M_{\text{Ni}}$, the SN energy $E_{\text{SN}}$, and the opacity $\kappa$; 
we set these to 0.1 $M_{\sun}$, $10^{51}$ erg, and 0.1 g cm$^{-2}$ respectively.  
Changing the ejected nickel mass makes very little difference to the light curve as far as $M_{\rm ej} \lesssim 1 \ M_\odot$, and 0.1 $M_{\sun}$ is typical of core collapse supernovae \citep[see, e.g.,][]{2011ApJ...741...97D}; 
the opacity is still uncertain, but the line opacity implies a value of 0.01-0.2 g cm$^{-2}$ at all times, so our value is not off by more than a factor of a few at any time \citep{Inserra_et_al_2013,2014MNRAS.438..318K};
the initial explosion energy is typical for ordinary supernovae and irrelevant since in the case of pulsar-driven SLSNe the rotation energy of the pulsar $\gtrsim 10^{52} \ \rm erg$, 
which is injected to the ejecta typically within $\sim$ a few days after the explosion, will dominate.
The light curves are fit to the model by eye.  
The fit does not include corrections for extinction, since the known $E(B-V)$ values are all $\lesssim$ 0.04 \citep{2017ApJ...835...64G}, although this does not include host galaxy extinction, which we expect to be low due to most SLSN residing in low metallicity dwarf galaxies~\citep{2014ApJ...787..138L}.  This means that the magnitudes may change by at most 0.15 mag, which will not make a large difference in the resulting parameters.  We base our fits on only the brightest 1.5 magnitudes due to the model's lack of reliability at later times
(see \cite{Kashiyama+16} for details). 

\begin{figure}%
    \includegraphics[width=0.5\textwidth]{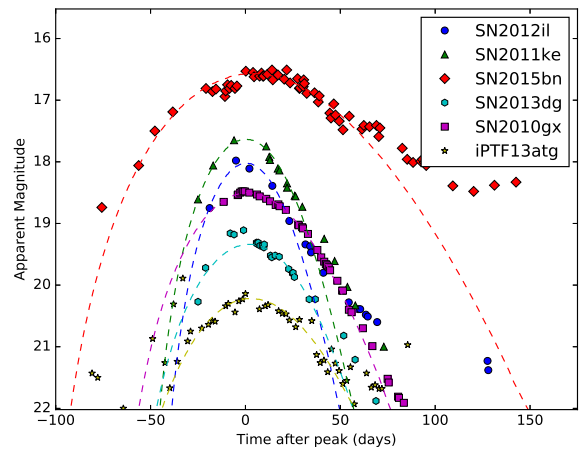}
    \caption{SN data (points) and modeled optical light curve (dashed lines) for each SN, using their $P$ = 1 ms parameter sets.}%
    \label{fig:snopmods}%
\end{figure}

We investigate a parameter range of $M_{\rm ej} \geq 1.0 \ M_{\rm ej}$ and $P \geq 1.0 \ \rm ms$. 
The former condition is reasonable for conventional core collapse explosions.  
The latter condition corresponds to the mass-shedding limit for NSs \citep{2016RvMP...88b1001W}.  
We found that all the sample SLSNe can be fit by the model with $P$ = 1.0 ms as shown in Fig. \ref{fig:snopmods}. 
Corresponding ($B_{13}, M_{\rm ej}$) are listed in the first and second column of Table. \ref{tbl:snprop},  
ranging from $B_{13} \sim 1-15$ and $M_{\rm ej} \sim 5-15 \ \rm M_\odot$. 

Notably, $(P, B_{13}, M_{\rm ej})$ suffer from parameter degeneracy; 
the SLSN light curves can be also fit with models with a slower initial spin.
The spin-down luminosity of the NS is sensitive to the initial spin period. 
For a slower spin, the ejecta acceleration is suppressed and the diffusion time becomes larger, roughly $t_{\rm dif} \propto P^{1/2} M_{\rm ej}^{1/2}$, and the peak time of the light curve becomes larger. 
Also, the peak SN luminosity is determined by the the spin-down luminosity at the diffusion time, roughly $L_{\rm sn} \propto P^{-1} B^{-2} M_{\rm ej}^{-1}$, and thus the SN becomes dimmer. 
The above effects of slower spin periods can be compensated by smaller values of the magnetic field and ejecta mass. 

\begin{figure}%
    \includegraphics[width=0.5\textwidth]{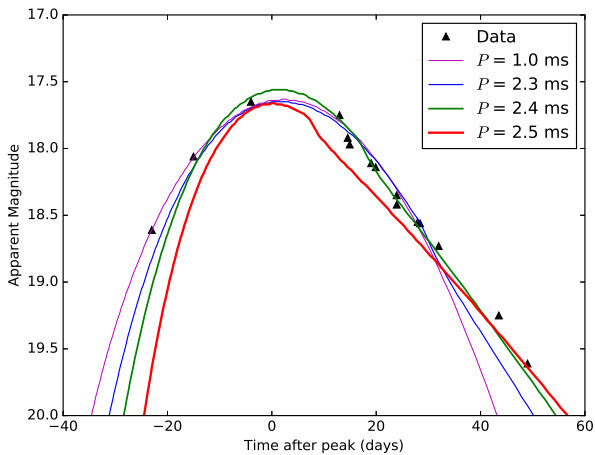}
    \caption{SN data (points) and modeled optical light curves for SN2011ke, using ($P$, $B_{13}$, $M_{\rm ej}$) = (1.0 ms, 7.5, 9.5 $M_{\sun}$), (2.3 ms, 3.7, 2 $M_{\sun}$), (2.4 ms, 2.9, 1.3 $M_{\sun}$), and (2.5 ms, 2.1, 0.8 $M_{\sun}$).  The $P$ = 1.0, 2.3, and 2.4 ms models are considered good fits, while the $P$ = 2.5 ms model is not wide enough at the peak and declines too slowly.  There are no parameter sets with $P$ = 2.5 ms that generate a model that fits the data, so 2.4 ms is considered to be $P_{\text{max}}$.}%
    \label{fig:pmax}%
\end{figure}

In Table. \ref{tbl:snparam}, we also show the slowest spin model that can fit each light curve. 
The slowest spin periods that give a reasonable fit to the data range from 1.1 to 4.1 ms depend on SNe;
a larger period would either bring the luminosity too low, or bring the field or ejecta mass low enough 
where the shape of the light curve would become inconsistent with the observed ones (Fig.~\ref{fig:pmax}). 

\begin{table*}
\caption{Model parameters of each SN that fits the light curve data.  Periods were investigated from 1.0 ms to $P_{\text{max}}$, with any period above $P_{\text{max}}$ either not having enough luminosity, having too slow a decline, or having a shape inconsistent with the observed data.  Data and fits for $P$ = 1.0 ms are shown in Fig.~\ref{fig:snopmods}.}
\begin{center}
\begin{tabular}{|c|c |c|c|c|c|} \hline
Name &  $B_{13}$ at 1 ms & $M_{\rm ej}$ ($M_{\sun}$) at 1 ms & $P_{\text{max}}$ (ms) &  $B_{13}$ at $P_{\text{max}}$ & $M_{\rm ej}$ ($M_{\sun}$) at $P_{\text{max}}$ \\ \hline 
iPTF13ajg & 1.6 & 5.0 & 1.1 & 1.3 & 3.5 \\
SN2012il & 8.0 & 7.0 & 2.4 & 3.0 & 1.0  \\
SN2013dg & 13.0 & 14.0 & 4.1 & 4.0 & 1.4  \\
SN2010gx & 4.5 & 10.0 & 1.6 & 3.5 & 3.5 \\
SN2011ke & 7.5 & 9.5 & 2.4 & 2.9 & 1.3 \\
SN2015bn & 2.1 & 17.0 & 1.4 & 1.0 & 5.0  \\ \hline 
\end{tabular}
\label{tbl:snparam}
\end{center}
\end{table*}

We should note that our fitting parameters for SN2012il, SN2011ke, and SN2010gx somewhat different from those used in \cite{Inserra_et_al_2013}. 
In particular, their best-fit models have a larger $B$ by a factor of $\lesssim 10$ and a larger $P$ by a factor of $\lesssim$ a few.    
An obvious reason for this apparent discrepancy is that we use the pulsar spin-down formula motivated by numerical simulations  \citep{gru05, spi06, tch+13}

\begin{equation}
L_{\text{em}} = \frac{\mu^2 \omega^4}{c^3}(1 + C\sin^2\chi_{\mu}),
\label{eqn:psrspindown}
\end{equation}

\noindent
where $\mu$ = $B_{\text{dip}}R^3/2$ is the magnetic moment, $\omega$ is the rotational angular frequency, $\chi_{\mu}$ is the angle between the magnetic and rotational axes, and $C$ $\sim$ 1 is a pre-factor; while \cite{Inserra_et_al_2013} used the classical dipole one.  The former gives a factor ${3(1+C\sin^2\chi_{\mu})/2\sin^2\chi_{\mu}} \sim 5$ larger spin-down luminosity than the latter for a fixed $(P, B)$~\citep[see][for a discussion]{Kashiyama+16}.  Also, the optical depth of the SN ejecta in our model is smaller than theirs by a factor of $\sim 1.5$ for a given $M_{\rm ej}$, $\kappa$, and the ejecta radius, which is simply caused by the different assumption on the SN ejecta profile; their ejecta has a homogeneous core surrounded by a homologous envelope with $\rho \propto v_{\rm ej}^{-10}$ where $v_{\rm ej}$ is the local velocity of the ejecta, while we only consider a homologous core with $\rho \propto v_{\rm ej}^{-1}$ \citep[see, e.g.,][]{kb10}.
These should be regarded as inherent uncertainties of the pulsar-driven SN model. 
In general, testing the pulsar-driven model only from optical/UV light curves is difficult since the SN light curve can be reproduced by the combination of different energy sources and other engines such as a black hole accretion disk are not excluded~\citep{Dexter_Kasen_13}. 
We do not investigate this problem further here. 
Instead we consistently calculate non-thermal counterparts in our framework, and examine the constraints placed by multi-wavelength observations. 

\section{Predictions for Radio Emission}\label{sec:radio}
In this section we calculate synchrotron radio counterparts of pulsar-driven SLSNe using the obtained model parameters in Sec. \ref{sec:SN}. We first calculate the early PWNe emission (Sec. \ref{sec:radioPWNe}) and then discuss radio emission from the ejecta forward shock (Sec. \ref{sec:ag}).  The non-thermal emission from PWNe is calculated as in previous papers~\citep[see][and references therein]{Gaensler_et_al_06,Tanaka_Takahara_10}, but we take into account not only dynamics of PWNe and SNe but also pair cascades and external attenuation based on \cite{2015ApJ...805...82M,MKM16}.

\subsection{Radio Emission from Embryonic Nebulae}\label{sec:radioPWNe}
In the course of the expansion, non-thermal PWNe emission starts to escape the SN ejecta. 
The timing of the escape sensitively depends on photon energy since the opacity depends on the energy, e.g., 
bound-free absorption for soft-X-rays, inelastic Compton scattering for hard-X-rays and MeV gamma-rays, Bethe-Heitler process for GeV gamma-rays, and photon-photon pair annihilation for higher energy photons.

The details of the PWNe spectrum depends on the injection history of electrons and positrons into the PWNe and the ionization state of the SN ejecta, both of which are uncertain for young pulsar-driven SNe. 
Here we assume an electron-positron injection spectrum motivated by Galactic PWNe such as the Crab PWNe~\citep[e.g.,][]{Tanaka_Takahara_10}, a broken power law with a peak Lorentz factor of $\gamma_{\rm e^{\pm},b} = 10^{5}$ and injection spectral indices of $q_1=1.5$ and $q_2=2.5$. 

Note that our results are not very sensitive to the spectral indices, since the resulting spectrum lies in the fast-cooling regime; the cooling time of electrons and positrons with $\gamma_{\rm e^{\pm},b}$ is much shorter than the dynamical time. The equipartition parameter for the magnetic field energy is assumed to be $\epsilon_B=0.01$ and the rest of the spin-down power is assumed to be used for the acceleration of electrons and positrons, i.e., $\epsilon_{\rm e} = 0.99$; this assumption is based on the results of detailed modeling of Galactic PWNe~\citep{Tanaka_Takahara_10, 2013MNRAS.429.2945T}. 
To take into account free-free absorption and the Razin effect in the SN ejecta, we also assume a singly ionized, oxygen-rich SN ejecta with an electron temperature of $T_{\rm e} = 10^4 \ {\rm K}$, a mean nuclear number of $\bar A=16$, and an effective atomic number of ${\bar Z}=4.5$ (taking into account the charge shielding effect); the assumption on the metal abundance is based on the results of previous nucleosynthesis studies~\citep[e.g.,][]{2002ApJ...565..405M}. The assumed ionized state can be maintained by the X-ray irradiation from the PWN, but the effect is difficult to calculate consistently.  
Although recent observations of PS1-14bj have found strong [O III] in the nebular spectra at $t \sim 0.5 \ \rm yr$ \citep{2016ApJ...831..144L}, which suggests a doubly-ionized state, the neutralization of the free electrons may proceed efficiently in the SN ejecta for $t \gtrsim 1 \ \rm yr$. In this sense, the assumption will overestimate free-free absorption, giving us a conservative estimate of the radio flux.

Other physical effects are calculated based on \cite{2015ApJ...805...82M,MKM16}.  The pulsar spin down, ejecta dynamics, and PWNe dynamics are handled in the same way as the~\cite{Kashiyama+16} model. Then, we solve a kinetic equation of electrons and positrons in the PWN, taking into account synchrotron and inverse Compton radiation, adiabatic cooling, and pair cascades. The injection spectrum of electrons and positrons to the PWN is assumed to be consistent with the Crab PWN~\citep{Tanaka_Takahara_10}. 

\begin{figure}%
    \includegraphics[width=\columnwidth]{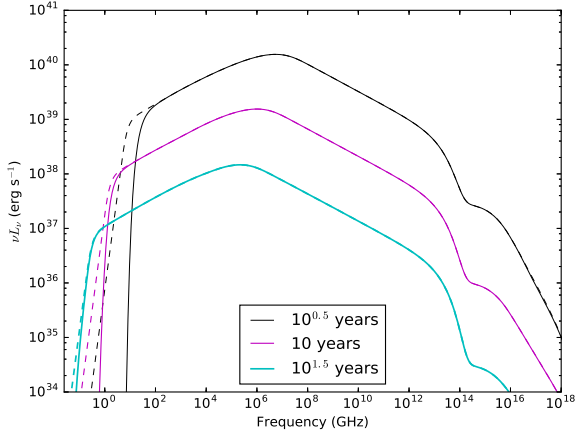}
    \caption{Broadband spectra from SN2011ke with $P = 1 \ \rm ms$ after $10^{0.5}$, 10, and $10^{1.5}$ years.  
    The solid lines take absorption processes for radio waves into account while the dashed lines do not.}%
    \label{fig:pwnspec}%
\end{figure}

The broadband spectra from SN2011ke with $P = 1 \ \rm ms$ at $10^{0.5}$, 10, and $10^{1.5}$ years are shown in Fig.~\ref{fig:pwnspec}. 
The spectrum is basically constituted from two components; the broken power-law spectrum up to $\nu \lesssim 10^{14} \ \rm GHz$ is from electron/positron synchrotron emission,  
while the bump at higher frequencies is due to inverse Compton scattering of the thermal photons in the SN ejecta.  
The peak frequency of the $\nu F_\nu$ spectrum corresponds to the synchrotron frequency of electrons/positrons with $\gamma_{\rm e^{\pm}, \rm peak}$. 
Overall, the spectrum becomes softer and the flux becomes smaller with time essentially due to the decline of the spindown luminosity and the adiabatic energy loss of the PWNe. 
The spectra from each SN is qualitatively similar, so only SN2011ke is shown.

\begin{figure}%
$\begin{array}{c}
    \includegraphics[width=0.5\textwidth]{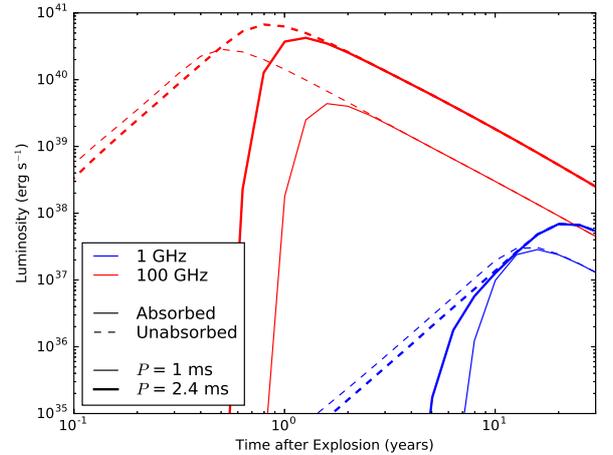} 
\end{array}$
    \caption{
    Intrinsic radio light curves from the pulsar wind nebula from SN2011ke at 1 GHz (blue) and 100 GHz (red).
    The thick and thin solid lines show the $P = 1 \ \rm ms$ and $P = P_{\rm max} = 2.4 \ \rm ms$ cases, respectively, including the absorption processes in the PWNe and SN ejecta. 
    The dashed lines are the unabsorbed light curves. 
    }%
    \label{fig:pwnin}%
\end{figure}

\begin{figure}%
$\begin{array}{c}
    \includegraphics[width=0.5\textwidth]{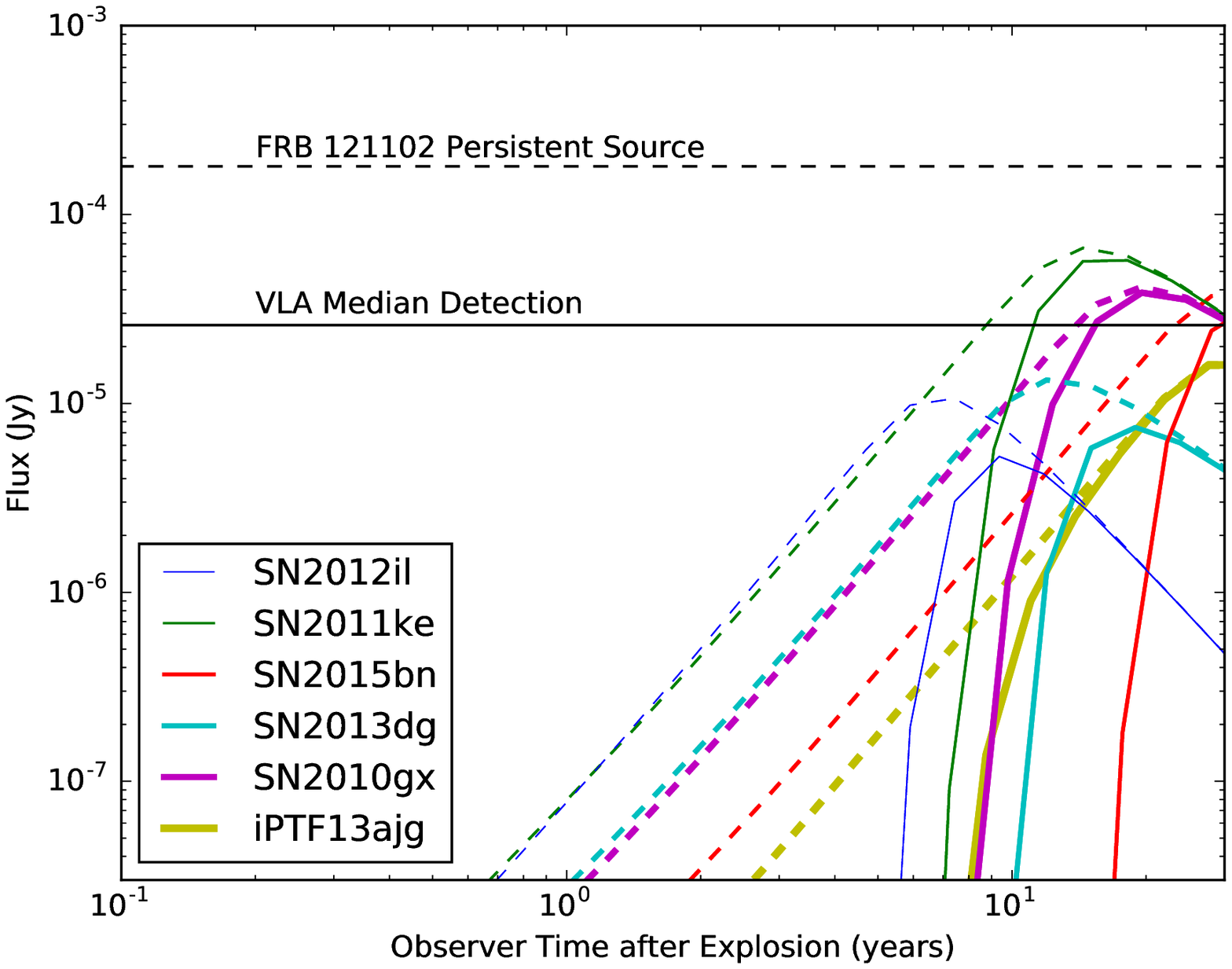} \\ 
    \includegraphics[width=0.5\textwidth]{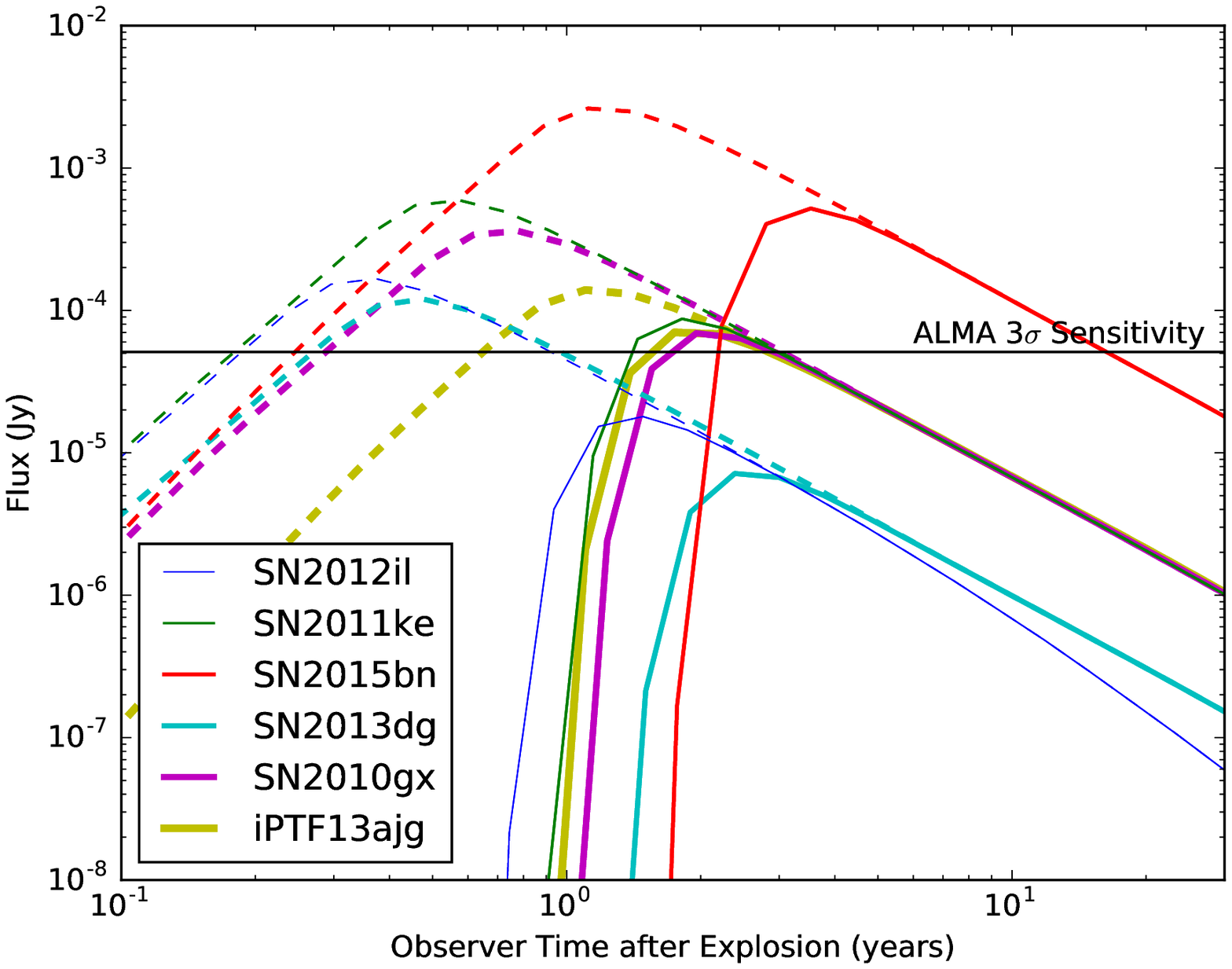}
\end{array}$
    \caption{Predicted observable light curves from the pulsar wind nebula from each SN at 1 GHz (above) and 100 GHz (below) using the $P$ = 1 ms parameters.  
    The solid lines indicate the light curve with maximum absorption, while the dashed lines indicate the light curve with no absorption.  
    The horizontal solid black lines indicate the median VLA detection (top) and the 3$\sigma$ detection limit from ALMA (bottom) taken from \citep{Chatterjee_et_al_17}.
    The dashed black line indicates the flux of the persistent source of FRB 121102 at $D_{\rm L} = 972 \ \rm Mpc$.  
    }%
    \label{fig:pwnob}%
\end{figure}

\begin{figure}%
$\begin{array}{c}
    \includegraphics[width=0.5\textwidth]{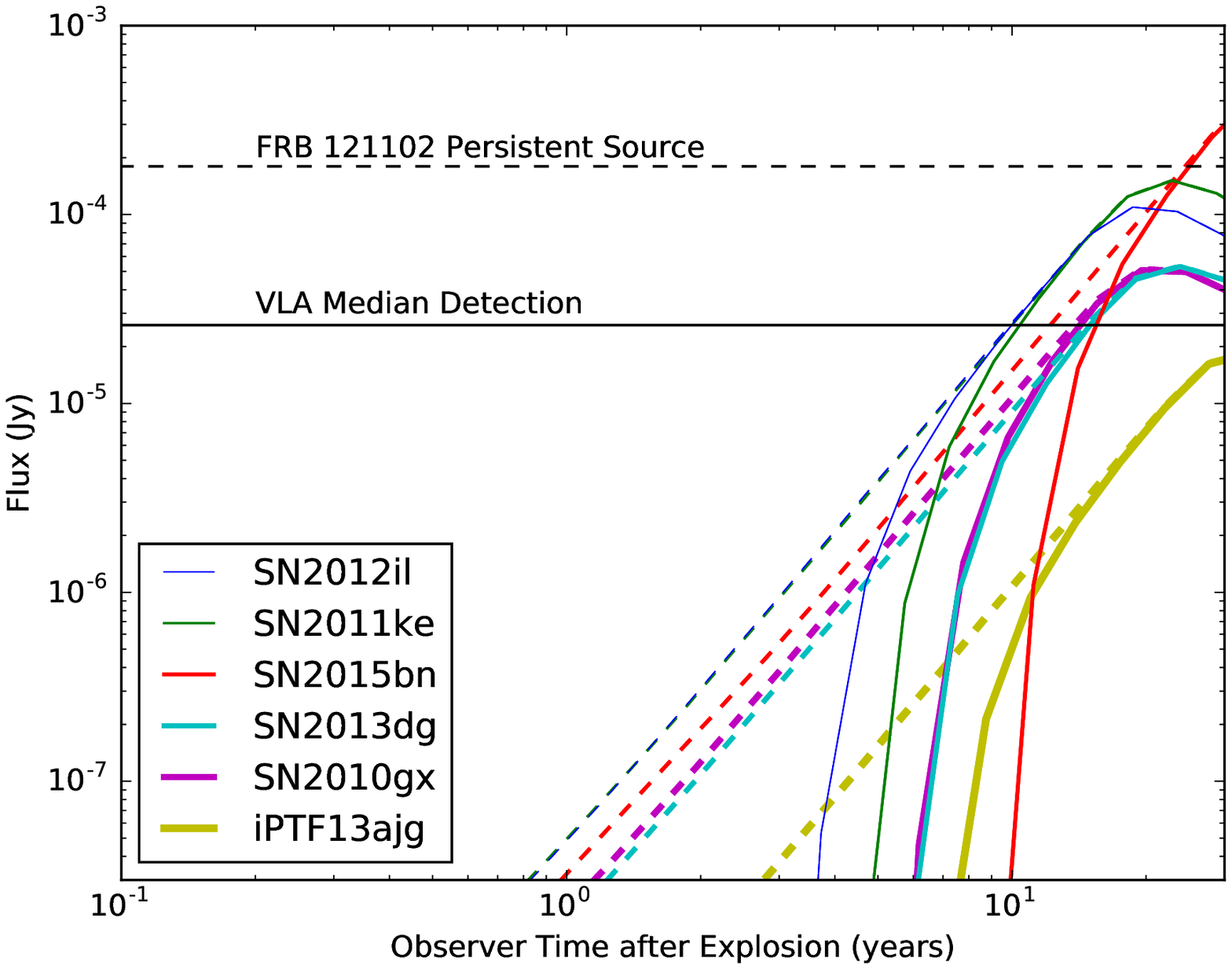} \\ 
    \includegraphics[width=0.5\textwidth]{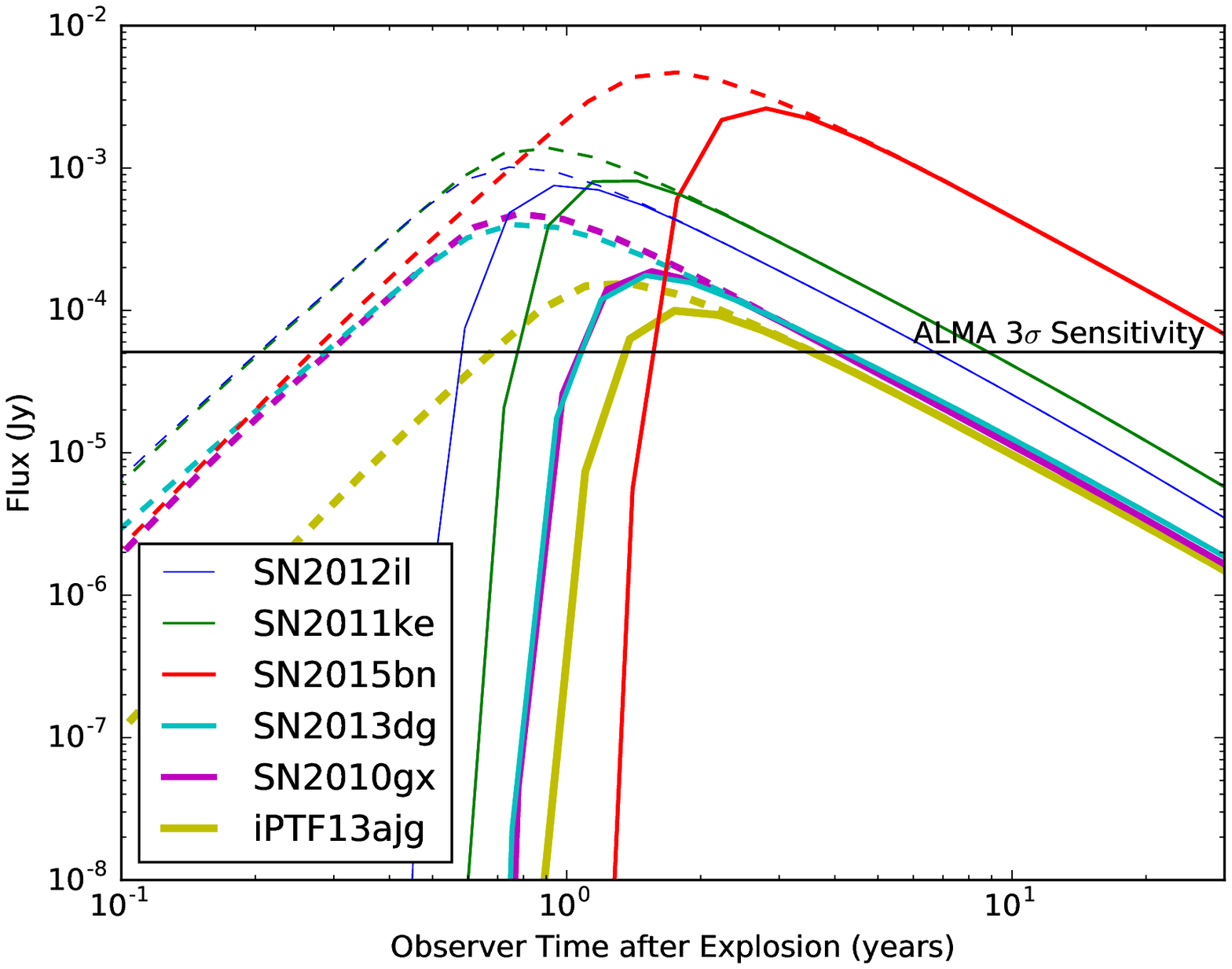}
\end{array}$
    \caption{The same as Figure~\ref{fig:pwnob}, but for the $P_{\text{max}}$ parameter sets from Table~\ref{tbl:snparam}.}%
    \label{fig:pwnobmax}%
\end{figure}

Hereafter, we focus on radio PWNe that radiate synchrotron emission of relativistic electrons and positrons that are accelerated in the nebulae.  The cooling time of high-energy electrons and positrons is shorter than the dynamical time of the system.  At later times, low-energy electrons and positrons do not have enough time to cool and they start to get accumulated in the PWNe. Then, the ``relic" electrons and positrons injected in the past can also contribute to the radio emission, which is consistently calculated in our model~\citep{MKM16}. 

Fig.~\ref{fig:pwnin} shows intrinsic light curves at the 100 GHz (red) and 1 GHz (blue) band for SN2011ke.
The solid lines include all the absorption processes while the dashed lines show the radio PWNe with no absorption in the SN ejecta (though the synchrotron-self absorption is included).  
The light-curve peak corresponds to the time when the radio frequency becomes comparable to the synchrotron self-absorption frequency, while the peak of the absorbed light curve is determined by free-free absorption. 
The absorption processes become irrelevant $\sim 1$ and $\sim 10$ yrs after the explosion for $\sim 100$ and $\sim 1$ GHz band, respectively. The decline after the peak is essentially due to the spin-down of the pulsar. 
In the same panel, we also show the dependence of the radio PWNe on the spin period. The solid thick and thin lines correspond to the fastest ($P = 1 \ \rm ms$) and slowest ($P = P_{\rm max} = 2.4 \ \rm ms$) spinning models allowed for SN2011ke, respectively.
As shown in Table \ref{tbl:snparam}, the slower rotating case requires a smaller magnetic field, 
which makes the spin-down time longer, and thus the spin-down luminosity can be kept relatively high for a longer time. 
Also the slower rotating case indicates a smaller ejecta mass, so the radio PWNe becomes transparent earlier. 
Consequently, the peak luminosity of the slower rotating case is brighter than the faster rotating case by a factor of $\lesssim10$. 

The light curves as would be observed from Earth in the 1 GHz band and 100 GHz band are shown in Fig.~\ref{fig:pwnob}. The $P$ = 1 ms parameter sets are adopted.  The solid lines indicate the light curve with our nominal absorption, while the dashed lines indicate the light curve with no absorption. For the 1 GHz band, VLA's 26 $\mu$Jy median flux density from 68 background sources around the persistent source is shown, and for the 100 GHz band the 51 $\mu$Jy 3$\sigma$ detection limit from ALMA.  
\footnote{Note that these limits actually come from the 3 GHz and 230 GHz band of VLA and ALMA, respectively.}
We find that radio emission from some SLSNe in our samples reach the VLA sensitivity in 10-20 years and the emission is detectable until 30 years after the explosion. If the absorption is suppressed by other effects such as mass shredding via Rayleigh-Taylor instabilities, the detection of radio sigals is possible at earlier times. 
In the 100 GHz band with  about our nominal absorption, several SLSNe in our sample have a peak submm flux that is close to the ALMA detection limit. 
If the absorption is suppressed and emission at the peak time is observed, 
non-thermal submm signals could be detected from 2-7 months until 1-2 years after the explosion.
Motivated by the possible connection between pulsar-driven SNe and FRBs~\citep{MKM16}, with the dashed black line, we show the 180 $\mu$Jy flux of the persistent source of FRB 121102 \citep{Chatterjee_et_al_17}.

In Fig.~\ref{fig:pwnobmax} we show radio light curves for each SN, using their $P_{\text{max}}$ parameter sets from Table~\ref{tbl:snparam}. 
At 1 GHz, all SNe show a flux peak at later times than in our fiducial cases, with a timescale of $\sim$ 20-30 years that is not heavily affected by absorption.  
All SNe are detectable by VLA except for iPTF13ajg, even with our nominal parameters including the absorption in the SN ejecta. 
Around its peak, SN2015bn is well above the FRB 121102 persistent source flux and SN2011ke is only slightly below. 
At 100 GHz, all SNe are detectable regardless of the absorption in the ejecta. 
Even with maximum absorption, the emission is detectable from 9 months to 2 years until 3-30 years after the explosion.  

\subsection{Radio Emission from Ejecta Forward Shocks}\label{sec:ag}
So far we have mainly considered radio emission from PWNe associated with pulsar-driven SLSNe. 
In addition to this component, we expect radio synchrotron emission from electrons accelerated at the SN forward shock. 
Here we estimate such a radio SN emission using a standard model~\citep[e.g.,][]{Chevalier_1998,Nakar_Piran_2011}.

With the model parameters in Table~\ref{tbl:snparam}, the pulsar central engine spins down within a few months after the explosion. 
Most of the initial rotation energy is converted into the kinetic energy of the SN ejecta, which is $E_{\rm K} \sim 2 \times 10^{52} \ {\rm erg} \ (P /1 \ {\rm ms})^{-2}$. 
The typical ejecta velocity is $v_{\rm ej} \approx (2E_{\rm K}/M_{\rm ej})^{1/2} \sim 0.06 \times c \ (P/1 \ {\rm ms})^{-1} (M_{\rm ej}/5 \ M_\odot)^{-1/2}$. 
In this case, the peak of the radio light curve with $\nu \gtrsim \rm GHz$ typically corresponds to the deceleration time of the SN ejecta, 
$t_{\rm dec} \sim 100 \ {\rm yr} \ (M_{\rm ej}/5 \ M_\odot)^{1/3} (n/1 \ {\rm cm^{-3}})^{-1/3}$, where $n$ is the number density of the interstellar medium. 
The peak flux can be estimated as $F_{\rm peak} \sim 65 \ {\rm \mu Jy} \ (\nu/1.5 \ {\rm GHz})^{-3/4}$ for a luminosity distance of $D_{\rm L} = 300 \ \rm Mpc$. 
In the above estimate, we assume that the power-law index of accelerated electrons as $p = 2.5$, the magnetic field amplification efficiency as $\epsilon_{\rm B} = 0.1$, 
and the electron acceleration efficiency as $\epsilon_{\rm e} = 0.1$; the flux decreases with smaller $\epsilon_{\rm B}$ and $\epsilon_{\rm e}$.
Before the peak, the flux evolves as $\propto t^3$. Note that the above parameter set for the radio emission from ejecta is optimistic; 
e.g., the ejecta kinetic energy is smaller for a slower rotating case. 
Comparing the above radio emission and the radio PWNe in Figs. \ref{fig:pwnob} and \ref{fig:pwnobmax}, we conclude that the latter likely dominates the former at least for a few decades after the explosion. 

Note that the radio emission from electrons accelerated the forward shock is more important in the presence of circumstellar material. 
In particular, Type-II SLSNe, which are believed to be interaction-powered, are expected to be strong radio sources in 1-10 year time scales~\citep{mur+14isn,Petropoulou_et_al_16}.

\section{Summary and Discussion}
In this work, we have investigated non-thermal radio signals from pulsar-driven SLSNe in a time window of a few decades after the explosion. 
We have found that the PWNe emission is likely to dominate the radio SN emission from the ejecta forward shock.  
In addition, we have shown that follow-up observations of SLSNe with ALMA and VLA in $\sim 1$ and $\sim 10$ yr time scales are promising for relatively nearby events at $D_{\rm L} \lesssim 1 \ \rm Gpc$. Successful detections would support the pulsar-driven model for SLSNe and will help us solve the degeneracy of model parameters. Even non-detections will be useful for us to constrain the magnetar paradigm for the diversity of different classes of stripped-SNe.  

In this work, we have assumed the simple one-dimensional model for the evolution of the SN ejecta and PWNe. The external absorption is taken into account, assuming a singly ionized state of the oxygen-rich SN ejecta. We note that our predictions are subject to non-negligible uncertainties. 
At lower frequencies, the synchrotron self-absorption process is relevant, which is taken into account in our calculations. Radio waves can also be absorbed by the free-free absorption and the Razin effect in the SN ejecta. These absorption processes are sensitive to the ionization state of the SN ejecta, and the ionization is caused by X-ray emission from PWNe and a reverse shock induced by the SN ejecta. With our nominal parameters, we could overestimate the radio absorption and thus underestimate the observable flux. For example, the neutralization of the SN ejecta may proceed efficiently before the ionization state decouples from the ejecta evolution a few years after the explosion~\citep[e.g.,][]{1984ApJ...287..282H}. 
Also, the ejecta being pushed by the strong magnetized wind will be subject to Rayleigh-Taylor instability and become ``patchy" and ``clumpy''~\citep[e.g.,][]{1992ApJ...395..540C,2001ApJ...563..806B, aro03,2005ApJ...619..839C,2016ApJ...832...73C,2017MNRAS.466.2633S}. 
Then, a fraction of the PWNe may be more easily observed due to its reduced optical depths, even if the average ionization degree of the ejecta is high.
If the wind bubble (that is surrounded by the PWNe mixed with the shocked SN ejecta) is largely blown out, the nebula radius rapidly increases, and the resulting spectra become similar to those of Galactic PWNe.  

Searches for radio counterparts of SLSNe are also of interest to test the connection between young NSs and FRBs. In this work, we have found that the radio emission from embryonic SLSN remnants with an age of a few decades is broadly consistent with a flux level of the persistent radio counterpart of FRB 121102. It is also intriguing that young NS scenarios for pulsar-driven SNe and FRBs predicted the existence of bright quasi-steady radio emission before the host galaxy of FRB 121102 was detected~\citep{MKM16}. 
On the other hand, in general, the PWNe emission itself does not have to be powered by the rotation energy. Indeed, \citep{Beloborodov_17} argued that the energy is supplied via the magnetic activity of a magnetar associated with the FRB.
Although this picture is different from ours, as long as the injected energy integrated over time is similar, both scenarios can lead to a similar prediction for nebular emission (although the two possibilities can be distinguished by a long-term follow-up observation of SLSNe with ALMA and VLA from $\sim 1$ to $10$ yrs), keeping the consistency with the FRB-SLSN connection. 

\section*{Acknowledgments}
We thank Akihiro Suzuki and Keiichi Maeda for discussion. 
C. M. B. O. acknowledges support by the MEXT University Recommended Scholarship.
K. K. acknowledges financial support from JST CREST. 
The work of K. M. is supported by NSF grant No. PHY-1620777.
While we were completing the draft, \cite{2017arXiv170102370M} appeared. Both papers share some points with the previous work \citep{MKM16}, regarding radio emission from pulsar-driven SNe and early PWNe.


\begin{thebibliography}{}
\makeatletter
\relax
\def\mn@urlcharsother{\let\do\@makeother \do\$\do\&\do\#\do\^\do\_\do\%\do\~}
\def\mn@doi{\begingroup\mn@urlcharsother \@ifnextchar [ {\mn@doi@}
  {\mn@doi@[]}}
\def\mn@doi@[#1]#2{\def\@tempa{#1}\ifx\@tempa\@empty \href
  {http://dx.doi.org/#2} {doi:#2}\else \href {http://dx.doi.org/#2} {#1}\fi
  \endgroup}
\def\mn@eprint#1#2{\mn@eprint@#1:#2::\@nil}
\def\mn@eprint@arXiv#1{\href {http://arxiv.org/abs/#1} {{\tt arXiv:#1}}}
\def\mn@eprint@dblp#1{\href {http://dblp.uni-trier.de/rec/bibtex/#1.xml}
  {dblp:#1}}
\def\mn@eprint@#1:#2:#3:#4\@nil{\def\@tempa {#1}\def\@tempb {#2}\def\@tempc
  {#3}\ifx \@tempc \@empty \let \@tempc \@tempb \let \@tempb \@tempa \fi \ifx
  \@tempb \@empty \def\@tempb {arXiv}\fi \@ifundefined
  {mn@eprint@\@tempb}{\@tempb:\@tempc}{\expandafter \expandafter \csname
  mn@eprint@\@tempb\endcsname \expandafter{\@tempc}}}

\bibitem[\protect\citeauthoryear{{Arnett}}{{Arnett}}{1982}]{1982ApJ...253..785A}
{Arnett} W.~D.,  1982, \mn@doi [\apj] {10.1086/159681}, \href
  {http://adsabs.harvard.edu/abs/1982ApJ...253..785A} {253, 785}

\bibitem[\protect\citeauthoryear{{Arons}}{{Arons}}{2003}]{aro03}
{Arons} J.,  2003, \mn@doi [\apj] {10.1086/374776}, \href
  {http://adsabs.harvard.edu/abs/2003ApJ...589..871A} {589, 871}

\bibitem[\protect\citeauthoryear{{Beloborodov}}{{Beloborodov}}{2017}]{Beloborodov_17}
{Beloborodov} A.~M.,  2017, preprint, \href
  {http://adsabs.harvard.edu/abs/2017arXiv170208644B} {} (\mn@eprint {arXiv}
  {1702.08644})

\bibitem[\protect\citeauthoryear{{Blondin}, {Chevalier}  \&
  {Frierson}}{{Blondin} et~al.}{2001}]{2001ApJ...563..806B}
{Blondin} J.~M.,  {Chevalier} R.~A.,   {Frierson} D.~M.,  2001, \mn@doi [\apj]
  {10.1086/324042}, \href {http://adsabs.harvard.edu/abs/2001ApJ...563..806B}
  {563, 806}

\bibitem[\protect\citeauthoryear{{Chatterjee} et~al.,}{{Chatterjee}
  et~al.}{2017}]{Chatterjee_et_al_17}
{Chatterjee} S.,  et~al., 2017, preprint, \href
  {http://adsabs.harvard.edu/abs/2017arXiv170101098C} {} (\mn@eprint {arXiv}
  {1701.01098})

\bibitem[\protect\citeauthoryear{{Chen}, {Woosley}  \& {Sukhbold}}{{Chen}
  et~al.}{2016}]{2016ApJ...832...73C}
{Chen} K.-J.,  {Woosley} S.~E.,   {Sukhbold} T.,  2016, \mn@doi [\apj]
  {10.3847/0004-637X/832/1/73}, \href
  {http://adsabs.harvard.edu/abs/2016ApJ...832...73C} {832, 73}

\bibitem[\protect\citeauthoryear{{Chevalier}}{{Chevalier}}{1998}]{Chevalier_1998}
{Chevalier} R.~A.,  1998, \apj, \href
  {http://adsabs.harvard.edu/abs/1998ApJ...499..810C} {499, 810}

\bibitem[\protect\citeauthoryear{{Chevalier}}{{Chevalier}}{2005}]{2005ApJ...619..839C}
{Chevalier} R.~A.,  2005, \mn@doi [\apj] {10.1086/426584}, \href
  {http://adsabs.harvard.edu/abs/2005ApJ...619..839C} {619, 839}

\bibitem[\protect\citeauthoryear{{Chevalier} \& {Fransson}}{{Chevalier} \&
  {Fransson}}{1992}]{1992ApJ...395..540C}
{Chevalier} R.~A.,  {Fransson} C.,  1992, \mn@doi [\apj] {10.1086/171674},
  \href {http://adsabs.harvard.edu/abs/1992ApJ...395..540C} {395, 540}

\bibitem[\protect\citeauthoryear{{Chevalier} \& {Irwin}}{{Chevalier} \&
  {Irwin}}{2011}]{Chevalier_Irwin_11}
{Chevalier} R.~A.,  {Irwin} C.~M.,  2011, \mn@doi [\apjl]
  {10.1088/2041-8205/729/1/L6}, \href
  {http://adsabs.harvard.edu/abs/2011ApJ...729L...6C} {729, L6}

\bibitem[\protect\citeauthoryear{{Dai}, {Wang}, {Wang}, {Wang}  \& {Yu}}{{Dai}
  et~al.}{2016}]{Dai_et_al_2016}
{Dai} Z.~G.,  {Wang} S.~Q.,  {Wang} J.~S.,  {Wang} L.~J.,   {Yu} Y.~W.,  2016,
  \mn@doi [\apj] {10.3847/0004-637X/817/2/132}, \href
  {http://adsabs.harvard.edu/abs/2016ApJ...817..132D} {817, 132}

\bibitem[\protect\citeauthoryear{{Dermer}, {Murase}  \& {Takami}}{{Dermer}
  et~al.}{2012}]{der+12}
{Dermer} C.~D.,  {Murase} K.,   {Takami} H.,  2012, \mn@doi [\apj]
  {10.1088/0004-637X/755/2/147}, \href
  {http://adsabs.harvard.edu/abs/2012ApJ...755..147D} {755, 147}

\bibitem[\protect\citeauthoryear{{Dexter} \& {Kasen}}{{Dexter} \&
  {Kasen}}{2013}]{Dexter_Kasen_13}
{Dexter} J.,  {Kasen} D.,  2013, \mn@doi [\apj] {10.1088/0004-637X/772/1/30},
  \href {http://adsabs.harvard.edu/abs/2013ApJ...772...30D} {772, 30}

\bibitem[\protect\citeauthoryear{{Drout} et~al.,}{{Drout}
  et~al.}{2011}]{2011ApJ...741...97D}
{Drout} M.~R.,  et~al., 2011, \mn@doi [\apj] {10.1088/0004-637X/741/2/97},
  \href {http://adsabs.harvard.edu/abs/2011ApJ...741...97D} {741, 97}

\bibitem[\protect\citeauthoryear{{Frail}, {Waxman}  \& {Kulkarni}}{{Frail}
  et~al.}{2000}]{2000ApJ...537..191F}
{Frail} D.~A.,  {Waxman} E.,   {Kulkarni} S.~R.,  2000, \mn@doi [\apj]
  {10.1086/309024}, \href {http://adsabs.harvard.edu/abs/2000ApJ...537..191F}
  {537, 191}

\bibitem[\protect\citeauthoryear{{Frail}, {Soderberg}, {Kulkarni}, {Berger},
  {Yost}, {Fox}  \& {Harrison}}{{Frail} et~al.}{2005}]{2005ApJ...619..994F}
{Frail} D.~A.,  {Soderberg} A.~M.,  {Kulkarni} S.~R.,  {Berger} E.,  {Yost} S.,
   {Fox} D.~W.,   {Harrison} F.~A.,  2005, \mn@doi [\apj] {10.1086/426680},
  \href {http://adsabs.harvard.edu/abs/2005ApJ...619..994F} {619, 994}

\bibitem[\protect\citeauthoryear{{Gaensler} \& {Slane}}{{Gaensler} \&
  {Slane}}{2006}]{Gaensler_et_al_06}
{Gaensler} B.~M.,  {Slane} P.~O.,  2006, \mn@doi [\araa]
  {10.1146/annurev.astro.44.051905.092528}, \href
  {http://adsabs.harvard.edu/abs/2006ARA%26A..44...17G} {44, 17}

\bibitem[\protect\citeauthoryear{{Gal-Yam}}{{Gal-Yam}}{2012}]{Gal-Yam12}
{Gal-Yam} A.,  2012, \mn@doi [Science] {10.1126/science.1203601}, \href
  {http://adsabs.harvard.edu/abs/2012Sci...337..927G} {337, 927}

\bibitem[\protect\citeauthoryear{{Greiner} et~al.,}{{Greiner}
  et~al.}{2015}]{Greiner_et_al_15}
{Greiner} J.,  et~al., 2015, \mn@doi [\nat] {10.1038/nature14579}, \href
  {http://adsabs.harvard.edu/abs/2015Natur.523..189G} {523, 189}

\bibitem[\protect\citeauthoryear{{Gruzinov}}{{Gruzinov}}{2005}]{gru05}
{Gruzinov} A.,  2005, \mn@doi [Physical Review Letters]
  {10.1103/PhysRevLett.94.021101}, \href
  {http://adsabs.harvard.edu/abs/2005PhRvL..94b1101G} {94, 021101}

\bibitem[\protect\citeauthoryear{{Guillochon}, {Parrent}, {Kelley}  \&
  {Margutti}}{{Guillochon} et~al.}{2017}]{2017ApJ...835...64G}
{Guillochon} J.,  {Parrent} J.,  {Kelley} L.~Z.,   {Margutti} R.,  2017,
  \mn@doi [\apj] {10.3847/1538-4357/835/1/64}, \href
  {http://adsabs.harvard.edu/abs/2017ApJ...835...64G} {835, 64}

\bibitem[\protect\citeauthoryear{{Hamilton} \& {Sarazin}}{{Hamilton} \&
  {Sarazin}}{1984}]{1984ApJ...287..282H}
{Hamilton} A.~J.~S.,  {Sarazin} C.~L.,  1984, \mn@doi [\apj] {10.1086/162687},
  \href {http://adsabs.harvard.edu/abs/1984ApJ...287..282H} {287, 282}

\bibitem[\protect\citeauthoryear{{Inserra} et~al.,}{{Inserra}
  et~al.}{2013}]{Inserra_et_al_2013}
{Inserra} C.,  et~al., 2013, \mn@doi [\apj] {10.1088/0004-637X/770/2/128},
  \href {http://adsabs.harvard.edu/abs/2013ApJ...770..128I} {770, 128}

\bibitem[\protect\citeauthoryear{{Kasen} \& {Bildsten}}{{Kasen} \&
  {Bildsten}}{2010}]{kb10}
{Kasen} D.,  {Bildsten} L.,  2010, \mn@doi [\apj]
  {10.1088/0004-637X/717/1/245}, \href
  {http://adsabs.harvard.edu/abs/2010ApJ...717..245K} {717, 245}

\bibitem[\protect\citeauthoryear{{Kashiyama} \& {Murase}}{{Kashiyama} \&
  {Murase}}{2017}]{2017ApJ...839L...3K}
{Kashiyama} K.,  {Murase} K.,  2017, \mn@doi [\apjl]
  {10.3847/2041-8213/aa68e1}, \href
  {http://adsabs.harvard.edu/abs/2017ApJ...839L...3K} {839, L3}

\bibitem[\protect\citeauthoryear{{Kashiyama}, {Murase}, {Bartos}, {Kiuchi}  \&
  {Margutti}}{{Kashiyama} et~al.}{2016}]{Kashiyama+16}
{Kashiyama} K.,  {Murase} K.,  {Bartos} I.,  {Kiuchi} K.,   {Margutti} R.,
  2016, \mn@doi [\apj] {10.3847/0004-637X/818/1/94}, \href
  {http://adsabs.harvard.edu/abs/2016ApJ...818...94K} {818, 94}

\bibitem[\protect\citeauthoryear{{Keane} \& {Kramer}}{{Keane} \&
  {Kramer}}{2008}]{kk08}
{Keane} E.~F.,  {Kramer} M.,  2008, \mn@doi [\mnras]
  {10.1111/j.1365-2966.2008.14045.x}, \href
  {http://adsabs.harvard.edu/abs/2008MNRAS.391.2009K} {391, 2009}

\bibitem[\protect\citeauthoryear{{Kleiser} \& {Kasen}}{{Kleiser} \&
  {Kasen}}{2014}]{2014MNRAS.438..318K}
{Kleiser} I.~K.~W.,  {Kasen} D.,  2014, \mn@doi [\mnras]
  {10.1093/mnras/stt2191}, \href
  {http://adsabs.harvard.edu/abs/2014MNRAS.438..318K} {438, 318}

\bibitem[\protect\citeauthoryear{{Kotera}, {Phinney}  \& {Olinto}}{{Kotera}
  et~al.}{2013}]{kot+13}
{Kotera} K.,  {Phinney} E.~S.,   {Olinto} A.~V.,  2013, \mn@doi [\mnras]
  {10.1093/mnras/stt680}, \href
  {http://adsabs.harvard.edu/abs/2013MNRAS.432.3228K} {432, 3228}

\bibitem[\protect\citeauthoryear{{Lunnan} et~al.,}{{Lunnan}
  et~al.}{2014}]{2014ApJ...787..138L}
{Lunnan} R.,  et~al., 2014, \mn@doi [\apj] {10.1088/0004-637X/787/2/138}, \href
  {http://adsabs.harvard.edu/abs/2014ApJ...787..138L} {787, 138}

\bibitem[\protect\citeauthoryear{{Lunnan} et~al.,}{{Lunnan}
  et~al.}{2016}]{2016ApJ...831..144L}
{Lunnan} R.,  et~al., 2016, \mn@doi [\apj] {10.3847/0004-637X/831/2/144}, \href
  {http://adsabs.harvard.edu/abs/2016ApJ...831..144L} {831, 144}

\bibitem[\protect\citeauthoryear{{Maeda}, {Nakamura}, {Nomoto}, {Mazzali},
  {Patat}  \& {Hachisu}}{{Maeda} et~al.}{2002}]{2002ApJ...565..405M}
{Maeda} K.,  {Nakamura} T.,  {Nomoto} K.,  {Mazzali} P.~A.,  {Patat} F.,
  {Hachisu} I.,  2002, \mn@doi [\apj] {10.1086/324487}, \href
  {http://adsabs.harvard.edu/abs/2002ApJ...565..405M} {565, 405}

\bibitem[\protect\citeauthoryear{{Marcote} et~al.,}{{Marcote}
  et~al.}{2017}]{Marcote_et_al_17}
{Marcote} B.,  et~al., 2017, \mn@doi [\apjl] {10.3847/2041-8213/834/2/L8},
  \href {http://adsabs.harvard.edu/abs/2017ApJ...834L...8M} {834, L8}

\bibitem[\protect\citeauthoryear{{Margutti} et~al.,}{{Margutti}
  et~al.}{2017}]{Margutti_et_al_17}
{Margutti} R.,  et~al., 2017, \mn@doi [\apj] {10.3847/1538-4357/836/1/25},
  \href {http://adsabs.harvard.edu/abs/2017ApJ...836...25M} {836, 25}

\bibitem[\protect\citeauthoryear{{Metzger}, {Vurm}, {Hasco{\"e}t}  \&
  {Beloborodov}}{{Metzger} et~al.}{2014}]{Metzger_et_al_2013}
{Metzger} B.~D.,  {Vurm} I.,  {Hasco{\"e}t} R.,   {Beloborodov} A.~M.,  2014,
  \mn@doi [\mnras] {10.1093/mnras/stt1922}, \href
  {http://adsabs.harvard.edu/abs/2014MNRAS.437..703M} {437, 703}

\bibitem[\protect\citeauthoryear{{Metzger}, {Margalit}, {Kasen}  \&
  {Quataert}}{{Metzger} et~al.}{2015}]{Metzger_et_al_15}
{Metzger} B.~D.,  {Margalit} B.,  {Kasen} D.,   {Quataert} E.,  2015, \mn@doi
  [\mnras] {10.1093/mnras/stv2224}, \href
  {http://adsabs.harvard.edu/abs/2015MNRAS.454.3311M} {454, 3311}

\bibitem[\protect\citeauthoryear{{Metzger}, {Berger}  \& {Margalit}}{{Metzger}
  et~al.}{2017}]{2017arXiv170102370M}
{Metzger} B.~D.,  {Berger} E.,   {Margalit} B.,  2017, preprint, \href
  {http://adsabs.harvard.edu/abs/2017arXiv170102370M} {} (\mn@eprint {arXiv}
  {1701.02370})

\bibitem[\protect\citeauthoryear{{Murase}, {Thompson}  \& {Ofek}}{{Murase}
  et~al.}{2014}]{mur+14isn}
{Murase} K.,  {Thompson} T.~A.,   {Ofek} E.~O.,  2014, \mn@doi [\mnras]
  {10.1093/mnras/stu384}, \href
  {http://adsabs.harvard.edu/abs/2014MNRAS.440.2528M} {440, 2528}

\bibitem[\protect\citeauthoryear{{Murase}, {Kashiyama}, {Kiuchi}  \&
  {Bartos}}{{Murase} et~al.}{2015}]{2015ApJ...805...82M}
{Murase} K.,  {Kashiyama} K.,  {Kiuchi} K.,   {Bartos} I.,  2015, \mn@doi
  [\apj] {10.1088/0004-637X/805/1/82}, \href
  {http://adsabs.harvard.edu/abs/2015ApJ...805...82M} {805, 82}

\bibitem[\protect\citeauthoryear{{Murase}, {Kashiyama}  \&
  {M{\'e}sz{\'a}ros}}{{Murase} et~al.}{2016}]{MKM16}
{Murase} K.,  {Kashiyama} K.,   {M{\'e}sz{\'a}ros} P.,  2016, \mn@doi [\mnras]
  {10.1093/mnras/stw1328}, \href
  {http://adsabs.harvard.edu/abs/2016MNRAS.461.1498M} {461, 1498}

\bibitem[\protect\citeauthoryear{{Nakar} \& {Piran}}{{Nakar} \&
  {Piran}}{2011}]{Nakar_Piran_2011}
{Nakar} E.,  {Piran} T.,  2011, \mn@doi [\nat] {10.1038/nature10365}, \href
  {http://adsabs.harvard.edu/abs/2011Natur.478...82N} {478, 82}

\bibitem[\protect\citeauthoryear{{Nicholl} et~al.,}{{Nicholl}
  et~al.}{2013}]{nic+13}
{Nicholl} M.,  et~al., 2013, \mn@doi [\nat] {10.1038/nature12569}, \href
  {http://adsabs.harvard.edu/abs/2013Natur.502..346N} {502, 346}

\bibitem[\protect\citeauthoryear{{Nicholl} et~al.,}{{Nicholl}
  et~al.}{2014}]{2014MNRAS.444.2096N}
{Nicholl} M.,  et~al., 2014, \mn@doi [\mnras] {10.1093/mnras/stu1579}, \href
  {http://adsabs.harvard.edu/abs/2014MNRAS.444.2096N} {444, 2096}

\bibitem[\protect\citeauthoryear{{Nicholl} et~al.,}{{Nicholl}
  et~al.}{2016}]{2016ApJ...828L..18N}
{Nicholl} M.,  et~al., 2016, \mn@doi [\apjl] {10.3847/2041-8205/828/2/L18},
  \href {http://adsabs.harvard.edu/abs/2016ApJ...828L..18N} {828, L18}

\bibitem[\protect\citeauthoryear{{Pastorello} et~al.,}{{Pastorello}
  et~al.}{2010a}]{2010ApJ...724L..16P}
{Pastorello} A.,  et~al., 2010a, \mn@doi [\apjl] {10.1088/2041-8205/724/1/L16},
  \href {http://adsabs.harvard.edu/abs/2010ApJ...724L..16P} {724, L16}

\bibitem[\protect\citeauthoryear{{Pastorello} et~al.,}{{Pastorello}
  et~al.}{2010b}]{Pastorello_et_al_2010}
{Pastorello} A.,  et~al., 2010b, \mn@doi [\apjl] {10.1088/2041-8205/724/1/L16},
  \href {http://adsabs.harvard.edu/abs/2010ApJ...724L..16P} {724, L16}

\bibitem[\protect\citeauthoryear{{Perna} \& {Stella}}{{Perna} \&
  {Stella}}{2004}]{Perna_Stella_2004}
{Perna} R.,  {Stella} L.,  2004, \mn@doi [\apj] {10.1086/423950}, \href
  {http://adsabs.harvard.edu/abs/2004ApJ...615..222P} {615, 222}

\bibitem[\protect\citeauthoryear{{Perna}, {Soria}, {Pooley}  \&
  {Stella}}{{Perna} et~al.}{2008}]{per+08}
{Perna} R.,  {Soria} R.,  {Pooley} D.,   {Stella} L.,  2008, \mn@doi [\mnras]
  {10.1111/j.1365-2966.2007.12821.x}, \href
  {http://adsabs.harvard.edu/abs/2008MNRAS.384.1638P} {384, 1638}

\bibitem[\protect\citeauthoryear{{Petropoulou}, {Kamble}  \&
  {Sironi}}{{Petropoulou} et~al.}{2016}]{Petropoulou_et_al_16}
{Petropoulou} M.,  {Kamble} A.,   {Sironi} L.,  2016, \mn@doi [\mnras]
  {10.1093/mnras/stw920}, \href
  {http://adsabs.harvard.edu/abs/2016MNRAS.460...44P} {460, 44}

\bibitem[\protect\citeauthoryear{{Quimby} et~al.,}{{Quimby}
  et~al.}{2011}]{qui+11}
{Quimby} R.~M.,  et~al., 2011, \mn@doi [\nat] {10.1038/nature10095}, \href
  {http://adsabs.harvard.edu/abs/2011Natur.474..487Q} {474, 487}

\bibitem[\protect\citeauthoryear{{Smith} \& {McCray}}{{Smith} \&
  {McCray}}{2007}]{Smith_McCray_07}
{Smith} N.,  {McCray} R.,  2007, \mn@doi [\apjl] {10.1086/524681}, \href
  {http://adsabs.harvard.edu/abs/2007ApJ...671L..17S} {671, L17}

\bibitem[\protect\citeauthoryear{{Spitkovsky}}{{Spitkovsky}}{2006}]{spi06}
{Spitkovsky} A.,  2006, \mn@doi [\apjl] {10.1086/507518}, \href
  {http://adsabs.harvard.edu/abs/2006ApJ...648L..51S} {648, L51}

\bibitem[\protect\citeauthoryear{{Suzuki} \& {Maeda}}{{Suzuki} \&
  {Maeda}}{2017}]{2017MNRAS.466.2633S}
{Suzuki} A.,  {Maeda} K.,  2017, \mn@doi [\mnras] {10.1093/mnras/stw3259},
  \href {http://adsabs.harvard.edu/abs/2017MNRAS.466.2633S} {466, 2633}

\bibitem[\protect\citeauthoryear{{Tanaka} \& {Takahara}}{{Tanaka} \&
  {Takahara}}{2010}]{Tanaka_Takahara_10}
{Tanaka} S.~J.,  {Takahara} F.,  2010, \mn@doi [\apj]
  {10.1088/0004-637X/715/2/1248}, \href
  {http://adsabs.harvard.edu/abs/2010ApJ...715.1248T} {715, 1248}

\bibitem[\protect\citeauthoryear{{Tanaka} \& {Takahara}}{{Tanaka} \&
  {Takahara}}{2013}]{2013MNRAS.429.2945T}
{Tanaka} S.~J.,  {Takahara} F.,  2013, \mn@doi [\mnras] {10.1093/mnras/sts528},
  \href {http://adsabs.harvard.edu/abs/2013MNRAS.429.2945T} {429, 2945}

\bibitem[\protect\citeauthoryear{{Tchekhovskoy}, {Spitkovsky}  \&
  {Li}}{{Tchekhovskoy} et~al.}{2013}]{tch+13}
{Tchekhovskoy} A.,  {Spitkovsky} A.,   {Li} J.~G.,  2013, \mn@doi [\mnras]
  {10.1093/mnrasl/slt076}, \href
  {http://adsabs.harvard.edu/abs/2013MNRAS.435L...1T} {435, L1}

\bibitem[\protect\citeauthoryear{{Tendulkar} et~al.,}{{Tendulkar}
  et~al.}{2017}]{Tendulkar_et_al_17}
{Tendulkar} S.~P.,  et~al., 2017, \mn@doi [\apjl] {10.3847/2041-8213/834/2/L7},
  \href {http://adsabs.harvard.edu/abs/2017ApJ...834L...7T} {834, L7}

\bibitem[\protect\citeauthoryear{{Thompson}, {Chang}  \& {Quataert}}{{Thompson}
  et~al.}{2004}]{Thompson_et_al_2004}
{Thompson} T.~A.,  {Chang} P.,   {Quataert} E.,  2004, \mn@doi [\apj]
  {10.1086/421969}, \href {http://adsabs.harvard.edu/abs/2004ApJ...611..380T}
  {611, 380}

\bibitem[\protect\citeauthoryear{{Vreeswijk} et~al.,}{{Vreeswijk}
  et~al.}{2014}]{2014ApJ...797...24V}
{Vreeswijk} P.~M.,  et~al., 2014, \mn@doi [\apj] {10.1088/0004-637X/797/1/24},
  \href {http://adsabs.harvard.edu/abs/2014ApJ...797...24V} {797, 24}

\bibitem[\protect\citeauthoryear{{Wang}, {Wang}, {Dai}  \& {Wu}}{{Wang}
  et~al.}{2015}]{Wang_et_al_2015}
{Wang} S.~Q.,  {Wang} L.~J.,  {Dai} Z.~G.,   {Wu} X.~F.,  2015, \mn@doi [\apj]
  {10.1088/0004-637X/807/2/147}, \href
  {http://adsabs.harvard.edu/abs/2015ApJ...807..147W} {807, 147}

\bibitem[\protect\citeauthoryear{{Watts} et~al.,}{{Watts}
  et~al.}{2016}]{2016RvMP...88b1001W}
{Watts} A.~L.,  et~al., 2016, \mn@doi [Reviews of Modern Physics]
  {10.1103/RevModPhys.88.021001}, \href
  {http://adsabs.harvard.edu/abs/2016RvMP...88b1001W} {88, 021001}

\bibitem[\protect\citeauthoryear{{Woosley}}{{Woosley}}{2010}]{Woosley_2010}
{Woosley} S.~E.,  2010, \mn@doi [\apjl] {10.1088/2041-8205/719/2/L204}, \href
  {http://adsabs.harvard.edu/abs/2010ApJ...719L.204W} {719, L204}

\makeatother
\end{thebibliography}

\bsp

\label{lastpage}

\end{document}